# Dual link failure survivability with recovery time constraint: A Parallel cross connection backup route recovery strategy

Dinesh Kumar, Rajiv Kumar, and Neeru Sharma
*Department of Electronics & Communication Engineering,*
*Jaypee University of Information Technology Waknaghat, Solan (HP) India*
* Corresponding author, Email:- dineshjuit@gmail.com, Mob: +91-8219718367, +91-9805584555

**Abstract**

In this paper, we proposed a fast recovery strategy for a dual link failure in elastic optical network. The elastic optical network is a promising solution to meet the next generation higher bandwidth demand. The survivability of high speed network is very crucial. As the network size increases the probability of the dual link failure and node failure also increases. Here, we proposed a parallel cross connection backup recovery strategy for dual link failure in the network. Proposed strategy shows lower bandwidth blocking probability (BBP), fast connection recovery, and bandwidth provisioning ratio when compared with the existing shared path protection (SPP) and dedicated path protection (DPP) approaches. Simulation is performed on ARPANET and COST239 topology networks.

**Keywords**

Dual link failure, Routing and spectrum assignment, Shared path protection, Dedicated path protection, Elastic optical network, and Frequency slot.

**Nomenclature**

| | |
|---|---|
| ADP | Aware differentiated protection |
| AFRO | After failure repair optimization |
| BVT | Bandwidth variable transponder |
| BBP | Bandwidth blocking probability |
| BPR | Bandwidth provisioning ratio |
| DLF | Dual link failure |
| DLP | Dedicated link protection |
| DPP | Dedicated path protection |
| DWDM | Dense wavelength division multiplexing |
| EON | Elastic optical network |
| FP | Fixed path |
| HAFA | Hybrid adaptive frequency assignment |
| HDVS | High definition video streaming |
| INCB-SPP | Intermediate node cross-connect backup for shared path protection |
| FS | Frequency slot |
| MFSB | Minimum free spectrum block consumption |
| NCU | Network capacity utilization |
| OPC | Online path computation |
| PBR | Primary backup route |
| RT | Recovery time |
| RTC | Recovery time constraint |
| RSA | Routing and spectrum assignment |
| SDN | Software defined network |
| SLP | Shared link protection |
| SPP | Shared path protection |
| SBPP | Shared backup path protection |
| SPP-MPR | Single path provisioning multipath recovery |
| TS | Tabu search |
| QoS | Quality of services |

## 1. Introduction

As reported by Cisco [1], threefold increase in global traffic in next five years up to 2022. The internet traffic grows rapidly globally at annual rate of 27% since 2017 to 2022. The Cisco forecasted the monthly mobile data rate by 2022 will be 77 exabytes and annually it will be one zettabyte [1] [2] and mobile occupied about 20% of the internet traffic by 2022. The bandwidth of optical fiber is very high, for next generation evaluation it requires extension in existing fixed grid dense wavelength division multiplexing (DWDM) network of 50 GHz or 100 GHz homogenous and fixed modulation format will not be able to cope with variable bandwidth demand [3], at low bit rate, which causes wastage of spectrum. Higher bit rate (400gb/s or 1Tb/s) demands are not met by existing fixed DWDM scheme [4]. Existing optical network needs to replace the opaque architectural to latest transparent network which have the energy efficient and considerable cost equipments [5].

The Elastic optical network (EON) with bandwidth and transponder flexibility has been viewed as spectrum efficient solution for higher bandwidth demand applications [6] [7]. Existing homogeneous optical network needs to replace the opaque architectural to latest transparent network which have the energy efficient and considerable cost equipments [5].

The bandwidth demand increases continuously due to the continuous growth in internet traffic [1] such as social network, cloud computing, multicasting, broadcasting, high definition video streaming (HDVS), and online gaming etc. All these applications required higher bandwidth for next generation high speed optical networks. EON replace the fixed channel DWDM with finer frequency slot (FS) channel width 6.25 GHz, 12.50 GHz, 25 GHz or 50 GHz [8] [9]. Bandwidth variable transponder (BVT) ranging from 10 Gbps to 200 Gbps [10]. The bandwidth requested by the users are integer multiple of these FS. Despite EON, the capacity of the fiber is limited by Shannon capacity limitation, the non linear spectral channel interaction is more important factor that limits the capacity of the optical fiber [11].

As the size of the optical network increases, correspondingly the probability of the single and dual link failure is also increases. In EON, the link failure causes vast data loss and huge disruption to the customers businesses and provides a poor quality of services (QoS) to the users. The failure in the network not only affect the quality of services but resulting a huge revenue loss to the network operator [12] and also cause huge congestion in the network.

The network survivability is the main issue in the EON. There are two main survivability schemes one is protection and other is restoration. In earlier scheme, the backup route is reserved in advance at the time of connection establishment and ensures to provide guaranteed 100% protection against any primary link failure in the optical network. The reserved FS not used until the failure happened in the network. In later scheme, the backup path is searched dynamically after the failure occurred in the network but not provided the guaranteed recovery of the network. The survivability issue in EON becomes more compelling due to spectrum continuity and contiguity constraint [13] [14]. The spectrum continuity constraint requires the same number of FS on each link, whereas the spectrum contiguity constraint follow the consecutive FS on each link [15]. In EON, both the working and backup route must need to gratify the spectrum continuity and contiguity constraints. These spectrum constraints cause more complexity while designing the EON as compared to fixed DWDM network. This paper presents a recovery strategy for dual link failure in EON [16]. Here, we consider that the two links are failing arbitrarily [17] and the recovery path provided are link disjoint and established at the time of connection setup.

This paper organized as follows: Section 2 provided the related work, the examples of dual link failure are explained in Section 3, the Notations and system model is presented in Section 4, the proposed recovery strategy for a dual link failure and existing schemes are explained in Section 5, simulation and comparison of the results are provided in Section 6, conclusion of the paper is presented in Section 7.

2. **Related Work**

In EON, the flexible sliced spectrum assignment is used for each request rather than fixed spectrum assignment in DWDM [18]. Most of previous work provides a survivability schemes for single link failure in EON. For EON different RSA model has been developed with or without considering the level of modulation [19][20]. The spectrum continuity and contiguity constraint are also considered for RSA in EON. The spectrally efficient and spatial flexible network is proposed in [21]. The survivability is the more critical issue in link or node failure in the optical network. The several network design has been proposed for the management of the spectrum in

EON [22], dual link failure [23], restoration [24], multipath protection [25], p-cycle design [26], mapping of virtual network design [27], multicast survivability [28] all are presented in literature with an objective of proficient utilization of the resources. The first and last fit recovery strategy has been proposed in [29]. The comparison between dedicated path protection (DPP) and shared backup path protection (SBPP) and integer linear programming (ILP) are addressed in [15][30]. The SPP is proposed in [22] and spectrum fragmentation and shared backup route is presented in [31]. The comparison of DPP with three strategies: 1) supporting graph based RSA, 2) first fit FS assignment, 3) and the routing path length is presented in [30]. The minimum free spectrum block consumption algorithm (MFSB) is presented in [32]. The rescaled failure probability aware algorithm by using ILP is discussed in [33]. The aware differentiated protection (ADP) algorithm agreement between the network operator and end users is explained in [34]. The performance of ADP, DPP and SPP is evaluated through open flow based software, the performance of ADP is better as compared to DPP and SPP. The shared path protection with correlated risk (SPPCR) is presented in [35]. The spectrum utilization and energy issue for fixed and flexible grid is addressed in [36]. For a connection request, correlated risk is used for route computations. Hybrid adaptive frequency assignment (HAFA/TS) and Tabu search for DPP and SPP is presented in [37]. After failure repair optimization (AFRO) scheme is addressed in [38]. Multilink failure survivability with dynamic load balancing is addressed in [39]. The different route selection polices for fixed path (FP) and online path computation (OPC) is presented in [40]. The multipath survivability with bandwidth clutching for SDN based EON is provided in [41]. The single path provisioning multipath recovery (SPP-MPR) with bitrate are presented in [42]. The higher resource sharing with multipath protection is explained in [25]. The survivable virtual optical networking which is also known as optimum shared protection mapping (OSPM) is provided in [27].

3. **Dual link failure**

In this Section, two cases for dual link failure are presented. In first case, different links with different primary routes fails simultaneously. And, in second case, links of primary and backup routes fails simultaneously. In Fig. 1 (a) we consider the first failure on primary route, $P_1$ (C-A), and simultaneously the second failure $P_2$ (B-A) occurred on primary route $P_2$. The backup of both these route are $B_1$ (C-D-A) and $B_2$ (B-D-A) for $P_1$ and $P_2$, respectively, which shared the same FS. Similarly in Fig. 1 (b) the failure occurred on primary route, $P_1$ (C-A), and simultaneously the backup route, $B_1$ of $P_1$ is also fails, for which the backup route $B_2$ (B-D-A) is provided. For the backup links we required additional spare capacity in the network and used only when it's necessary. In our proposed INCB-SPP, the primary backup route (PBR) always exists.

## 4. Notations Used

G (V, L, F): Consider a network topology with $|V|$ nodes and $|L|$ links. Each link $l \in L$ and have its beginning node $s(l)$ and receiving node $r(l)$.

| | |
|---|---|
| $v$ | Nodes in the network topology, $(v_1, v_2, v_3, .....v_n$ be the set of the nodes in the network) $v \in V$ |
| $l$ | Links in the network topology, $(l_1, l_2, l_3, ......l_n$ be the set of links in the network) $l \in L$ |
| $f$ | Set of FS on each link, $(f_1, f_2, f_3, .......f_n)$ indexes as $f \in F$. |
| $d$ | Set of traffic request, each request has its origin at $s(d)$ and $r(d)$ is the receiving end $d \in D$ |
| $w$ | Working / primary path in the network $w \in W$ |
| $r_b$ | Reserved backup path in the network $r_b \in R_b$ |
| $\ell^f$ | $f^{th}$ frequency slot on link $\ell$ |
| $\ell_w^f$ | $f^{th}$ frequency slot of $w^{th}$ working route on link $\ell$ |
| $l_{r_b}^f$ | $f^{th}$ frequency slot of $r_b^{th}$ backup route on link $\ell$ |
| $RT_r$ | Recovery time of request $r$, $\forall r \in R$ |
| $R_{tc}$ | Recovery time constraint |

Evolution of different network parameters is as follows:

- Failure detection time, is the time taken by the node to detect the failure, we consider the failure detection time, $f_d$ is 10μs [15].
- Message processing ($M_p$) time at the node is 10μs.
- Signal propagation delay ($s_{pd}$) on each link is 400μs for each 85km.
- Cross connection, is the connection establishment time ($c_x$), we consider the cross connection time is 2ms.
- Number of links ($b_l$) in the backup light path.
- $p_t$ be the propagation time of the message.
- $n$, be the Number of nodes and $M_a$ be the message acknowledgement time from destination to source and $l$ be the length of the backup light path.

**Network constraint**

Different network constraints have been considered while designing the survivable network.

**3.1 Maximum capacity of the link**

The number of FS is equal to or less than the number of FS available on that link.

$$\sum_{\forall f \in F} l^f, \qquad \forall l \in L \tag{1}$$

## 3.2 Spectrum continuity and contiguity constraint for primary and backup route

For continuity constraint, this constraint require same number of FS to each link

(i) For primary path

$$l^f_{w_i} - l^f_{w_k} = 0, \forall i \neq k, \forall f \,\varepsilon\, F, \forall w \,\varepsilon\, W, \forall l \,\varepsilon\, L \tag{2}$$

(ii) For backup path (Reserved path)

$$l^f_{r_{b_i}} - l^f_{r_{b_k}} = 0, \forall i \neq k, \forall f \,\varepsilon\, F, \forall r_b \,\varepsilon\, R_b, \forall l \,\varepsilon\, L \tag{3}$$

For contiguity constraint

In contiguity constraint, we consider the number of FS request arrives is *FR*.

(i) For primary path

$$l^{f_i}_w - l^{f_k}_w = FR - 1, \forall i \neq k, \forall f \,\varepsilon\, F, \forall w \,\varepsilon\, W, \forall l \,\varepsilon\, L \tag{4}$$

(ii) For backup path (Reserved path)

$$l^{f_i}_{r_b} - l^{f_k}_{r_b} = FR - 1, \forall i \neq k, f \,\varepsilon\, F, \forall r_b \,\varepsilon\, R_b, \forall l \,\varepsilon\, L \tag{5}$$

## 4.3 Recovery time constraint (RTC)

RTC is the maximum connection setup time of the network after the failure. $T_{rcs}$ be the recovery time of the connection setup and $R_{tc}$ is the recovery time constraint.

$$T_{rcs} \leq R_{tc} \tag{6}$$

## 5. Proposed and Existing Strategies

Here in this Section, we explained the existing survivability schemes i.e. SPP, and DPP, and then our proposed recovery scheme that is intermediate node cross-connects backup route for shared path protection (INCB-SPP). The different failure scenarios of failure are presented in Figure 1 (a) and (b).

### 5.1 Shared path protection (SPP)

In this scheme, the spare capacity is shared with other light paths. In SPP, the backup path is pre-decided; the FS is reserved in advance but not used, until the failure occurred in the network. When a working link fails, the whole traffic is diverted to reserved backup path. For a recovery of a single link failure, the entire working path in the network must share the same unused capacity that is link disjoint to each other. SPP requires less than the DPP. In SPP the cross connection at the nodes increases the connection recovery time. The total connection recovery time for this scheme is given as

$$RT_{spp} = 2 \times n \times (M_p + M_a) + n \times c_x + 2 \times l \times p_t \tag{7}$$

## 5.2 Dedicated path protection (DPP)

In this scheme, the DPP is a 1+1 scheme, intermediate node of backup route is already cross-connect at the time of connection setup. In DPP [30] no cross connection are used for the backup route, hence in this case the recovery time is lower than the SPP. The total connection recovery time DPP is as under

$$RT_{dpp} = 2 \times n \times (M_p + M_a) + 2 \times l \times p_t \tag{8}$$

## 5.3 Proposed Recovery scheme

In proposed scheme, we have decided an intermediate node where signal exchanged information by parallel reception of connection setup messages by source and destination nodes of the failed primary route. If the recovery time is less than the pre-decided time constraints, then connection is accepted, otherwise, it is rejected.

$$RT_t = T_{cs} + T_{ack} \tag{9}$$

Here, we consider $RT_t$ be the total recovery time, $T_{cs}$ and $T_{ack}$ are the total connection setup time and the total acknowledgement time for connection recovery.

## 6. Results and Analysis

In this paper, we consider three existing topologies as shown in Fig. 2 (a) and (b) that is ARPANET (US network) and COST239 and evaluate their performance for backup route by randomly generated source (s)-destination (d) connection request by using *N(N-1)/2*, where *N* be the number of nodes and simulate in MATLAB 2018 on i5, Intel core processor with 8GB RAM. The FS for heterogeneously random connections are generated for ARPANET and COST239. The three network parameter performance is evaluated for ARPANET and COST239 is as under.

## 6.1 Bandwidth blocking probability (BBP)

Bandwidth blocking probability is defined as the ratio of the total number of request rejected to the total number of request demanded in the network. The BBP is directly proportional to the total number of request available in the network. When the number of arriving connection request are lesser and the network capacity is larger, than the connection request acceptance ratio is more. As the number of connection request increases the rejection of connection request also increases and the acceptance ratio decreases. Fig. 3 (a) and (b) presents the BBP for all three survival strategies for ARPANET i.e. DPP, SPP and for Intermediate node cross-connect backup for shared path protection (INCB-SPP). The BBP of our proposed scheme is very less than DPP and SPP it goes on increasing as the connection request increases. The average of BBP for ARPANET is 0.7072, 0.4041 and 0.3828 for DPP, SPP and INCB-SPP respectively and for COST239 is 0.3947, 0.0455 and 0.0154 for DPP, SPP and INCB-SPP respectively.

**6.2 Recovery time (RT)**

Recovery time is the time at which the recovery process in the network started and the connection setup message received. RT of INCB-SPP must be less than SPP and DPP. The recovery time constraint (RTC) has been introduced for achieving the fast recovery of the failure in the network. The RTC for ARPANET and COST239 set 45 ms and 21 ms, respectively. The RT for our INCB-SPP strategy is very less and going on decreasing as the number of connection increases. The average RT for ARPANET is 8715μs, 16338μs and 4683μs for DPP, SPP and for INCB-SPP and for COST239 is 8201μs, 12474μs and 3799μs for DPP, SPP and INCB-SPP respectively.

**6.3 Bandwidth provisioning ratio (BPR)**

BPR is the ratio of the total FS used in the network to the total FS request accepted in the network. The average value of BPR for ARPANET is 6.0223, 2.8712 and 2.7161 for DPP, SPP and INCB-SPP and for COST239 is 3.5045, 1.7549 and 1.7149 for DPP, SPP and for INCB-SPP respectively. In our proposed INCB-SPP and SPP the sharing of network resources increases with the increase in the number of request, hence, the BPR in either case for both the topologies decreases with the increase in connection request as given in Fig. 5 (a) and (b). In case of DPP only the shorter routes accepted, for longer route the sharing of resources are unavailable.

**7. Conclusion**

In this paper, we proposed a realistic protection framework for a dual link failure in EON. Despite providing a 1+1 conventional protection (DPP) and SPP, we introduced an INCB-SPP recovery scheme for a primary link failure and backup route failure in EON. The proposed strategy not only provides the fast connection recovery but also utilizes network resources optimally. The maximum survivability for a dual link failure can be achieved by providing small additional spare capacity in the optical network.

**References**

[1]  Cisco, S. Jose, Cisco visual networking index (VNI) global mobile data traffic forecast update, 2017-2022 white paper, Ca, Usa. (2019) 3–5. http://www.gsma.com/spectrum/wp-content/uploads/2013/03/Cisco_VNI-

global-mobile-data-traffic-forecast-update.pdf.

[2] N.H. Bao, S. Sahoo, M. Kuang, Z.Z. Zhang, Adaptive path splitting based survivable virtual network embedding in elastic optical networks, Opt. Fiber Technol. 54 (2020) 102084. https://doi.org/10.1016/j.yofte.2019.102084.

[3] D.S. Yadav, S. Rana, S. Prakash, Optical Fiber Technology Hybrid connection algorithm : A strategy for efficient restoration in WDM optical networks, Opt. Fiber Technol. 16 (2010) 90–99. https://doi.org/10.1016/j.yofte.2009.12.002.

[4] A.D. Ellis, N. Mac Suibhne, D. Saad, D.N. Payne, Communication networks beyond the capacity crunch, Philos. Trans. R. Soc. A Math. Phys. Eng. Sci. 374 (2016). https://doi.org/10.1098/rsta.2015.0191.

[5] O. Gerstel, Elastic Optical Networking : A New Dawn for the Optical Layer ?, IEEE Commun. Mag. 50 (2012) s12–s20. https://doi.org/10.1109/MCOM.2012.6146481.

[6] B.A.A.M. Saleh, J.M. Simmons, All-Optical Networking V Evolution , Benefits , Challenges , and Future Vision, (2012).

[7] A. Dupas, P. Layec, E. Dutisseuil, S. Belotti, S. Bigo, E.H. Salas, G. Zervas, D. Simeonidou, Elastic optical interface with variable baudrate: Architecture and proof-of-concept, J. Opt. Commun. Netw. 9 (2017) A170–A175. https://doi.org/10.1364/JOCN.9.00A170.

[8] D.S. Yadav, S. Babu, B.S. Manoj, Quasi Path Restoration: A post-failure recovery scheme over pre-allocated backup resource for elastic optical networks, Opt. Fiber Technol. 41 (2018) 139–154. https://doi.org/10.1016/j.yofte.2018.01.011.

[9] G. F, C. Bbc, * C @ 9G 5B8 9B9 : = HG C : @ 5GH = 7 ' DH = 75 @ & 9HKCF ? G = B 9MCB8 6, (2016) 634–636.

[10] M. Klinkowski, K. Walkowiak, On the advantages of elastic optical networks for provisioning of cloud computing traffic, IEEE Netw. 27 (2013) 44–51. https://doi.org/10.1109/MNET.2013.6678926.

[11] P.P. Mitra, J.B. Stark, Nonlinear limits to the information capacity of optical fibre communications, Nature. 411 (2001) 1027–1030. https://doi.org/10.1038/35082518.

[12] Y. Ran, Considerations and Suggestions on Improvement of Communication Network Disaster Countermeasures after the Wenchuan Earthquake, (2011) 44–47.

[13] D. Batham, D. Singh, S. Prakash, Optical Fiber Technology Least loaded and route fragmentation aware RSA strategies for elastic optical networks, Opt. Fiber Technol. 39 (2017) 95–108. https://doi.org/10.1016/j.yofte.2017.10.003.

[14] D. Batham, D. Singh, Y. Shashi, Survivability using traffic balancing and backup resource reservation in multi - domain optical networks, (2018) 1–22. https://doi.org/10.1002/dac.3786.

[15] C. Wang, G. Shen, S.K. Bose, Distance Adaptive Dynamic Routing and Spectrum Allocation in Elastic Optical Networks with Shared Backup Path Protection, J. Light. Technol. 33 (2015) 2955–2964. https://doi.org/10.1109/JLT.2015.2421506.

[16] M. Sivakumar, K.M. Sivalingam, On surviving dual-link failures in path protected optical WDM mesh networks, Opt. Switch. Netw. 3 (2006) 71–88. https://doi.org/10.1016/j.osn.2006.04.004.

[17] P. Sasithong, L.Q. Quynh, P. Saengudomlert, P. Vanichchanunt, N.H. Hai, L. Wuttisittikulkij, Maximizing

double-link failure recovery of over-dimensioned optical mesh networks, Opt. Switch. Netw. 36 (2020) 100541. https://doi.org/10.1016/j.osn.2019.100541.

[18] M. Jinno, H. Takara, B. Kozicki, Y. Tsukishima, Y. Sone, S. Matsuoka, Spectrum-efficient and scalable elastic optical path network: Architecture, benefits, and enabling technologies, IEEE Commun. Mag. 47 (2009) 66–73. https://doi.org/10.1109/MCOM.2009.5307468.

[19] L. Gong, X. Zhou, X. Liu, W. Zhao, W. Lu, Z. Zhu, Efficient resource allocation for all-optical multicasting over spectrum-sliced elastic optical networks, J. Opt. Commun. Netw. 5 (2013) 836–847. https://doi.org/10.1364/JOCN.5.000836.

[20] M. Klinkowski, K. Walkowiak, Routing and spectrum assignment in spectrum sliced elastic optical path network, IEEE Commun. Lett. 15 (2011) 884–886. https://doi.org/10.1109/LCOMM.2011.060811.110281.

[21] D. Klonidis, F. Cugini, O. Gerstel, M. Jinno, V. Lopez, E. Palkopoulou, M. Sekiya, D. Siracusa, G. Thouénon, C. Betoule, Spectrally and spatially flexible optical network planning and operations, IEEE Commun. Mag. 53 (2015) 69–78. https://doi.org/10.1109/MCOM.2015.7045393.

[22] H. Liu, R. Li, Y. Chen, X. Wang, Resource efficiency improved approach for shared path protection in EONs, Photonic Netw. Commun. 33 (2017) 19–25. https://doi.org/10.1007/s11107-016-0612-9.

[23] B. Chen, J. Zhang, Y. Zhao, C. Lv, W. Zhang, S. Huang, X. Zhang, Optical Fiber Technology Multi-link failure restoration with dynamic load balancing in spectrum-elastic optical path networks, Opt. Fiber Technol. 18 (2012) 21–28. https://doi.org/10.1016/j.yofte.2011.10.002.

[24] L. Ruan, Y. Zheng, Dynamic survivable multipath routing and spectrum allocation in OFDM-based flexible optical networks, J. Opt. Commun. Netw. 6 (2014) 77–85. https://doi.org/10.1364/JOCN.6.000077.

[25] D.S. Yadav, A. Chakraborty, B.S. Manoj, Optical Fiber Technology A Multi-Backup Path Protection scheme for survivability in Elastic Optical Networks, Opt. Fiber Technol. 30 (2016) 167–175. https://doi.org/10.1016/j.yofte.2016.05.003.

[26] F. Ji, X. Chen, W. Lu, J.J.P.C. Rodrigues, Z. Zhu, Dynamic p-cycle protection in spectrum-sliced elastic optical networks, J. Light. Technol. 32 (2014) 1190–1199. https://doi.org/10.1109/JLT.2014.2300337.

[27] H. Yang, X. Zhu, W. Bai, Y. Zhao, J. Zhang, Z. Liu, Z. Zhou, Q. Ou, Survivable VON mapping with ambiguity similitude for differentiable maximum shared capacity in elastic optical networks, Opt. Fiber Technol. 31 (2016) 138–146. https://doi.org/10.1016/j.yofte.2016.07.002.

[28] A. Cai, Z. Fan, K. Xu, M. Zukerman, C.K. Chan, Elastic versus WDM networks with dedicated multicast protection, J. Opt. Soc. Am. 9 (2017) 921–933. https://doi.org/10.1364/JOCN.9.000921.

[29] A. Tarhan, Shared Path Protection for Distance Adaptive Elastic Optical Networks under Dynamic Traffic, IEEE Conf. (2013) 62–67.

[30] M. Klinkowski, K. Walkowiak, Offline RSA algorithms for elastic optical networks with dedicated path protection consideration, Int. Congr. Ultra Mod. Telecommun. Control Syst. Work. (2012) 670–676. https://doi.org/10.1109/ICUMT.2012.6459751.

[31] H. Liu, M. Zhang, P. Yi, Y. Chen, Shared path protection through reconstructing sharable bandwidth based on spectrum segmentation for elastic optical networks, Opt. Fiber Technol. 32 (2016) 88–95. https://doi.org/10.1016/j.yofte.2016.10.001.

[32] B. Chen, J. Zhang, Y. Zhao, H. Chen, S. Huang, W. Gu, J.P. Jue, Minimized spectral resource consumption with rescaled failure probability constraint in flexible bandwidth optical networks, J. Opt. Soc. Am. 5 (2013) 980–993. https://doi.org/10.1364/ofc.2013.otu3a.3.

[33] B. Chen, J. Zhang, Y. Zhao, J.P. Jue, J. Liu, S. Huang, W. Gu, Minimum Spectrum Block Consumption for Shared-Path Protection with Joint Failure Probability in Flexible Bandwidth Optical Networks, Opt. Switch. Netw. 13 (2014) 49–62. https://doi.org/10.1016/j.osn.2014.01.001.

[34] X. Chen, M. Tornatore, S. Zhu, F. Ji, W. Zhou, C. Chen, D. Hu, L. Jiang, Z. Zhu, Flexible availability-aware differentiated protection in software-defined elastic optical networks, J. Light. Technol. 33 (2015) 3872–3882. https://doi.org/10.1109/JLT.2015.2456152.

[35] J. Zhang, C. Lv, Y. Zhao, B. Chen, X. Li, S. Huang, W. Gu, A novel shared-path protection algorithm with correlated risk against multiple failures in flexible bandwidth optical networks, Opt. Fiber Technol. 18 (2012) 532–540. https://doi.org/10.1016/j.yofte.2012.09.002.

[36] J. López Vizcaíno, P. Soto, Y. Ye, P.M. Krummrich, Differentiated quality of protection: An energy- and spectral-efficient resilience scheme for survivable static and dynamic optical transport networks with fixed- and flexible-grid, Opt. Switch. Netw. 19 (2016) 78–96. https://doi.org/10.1016/j.osn.2015.03.006.

[37] K. Walkowiak, M. Klinkowski, B. Rabiega, R. Goścień, Routing and spectrum allocation algorithms for elastic optical networks with dedicated path protection, Opt. Switch. Netw. 13 (2014) 63–75. https://doi.org/10.1016/j.osn.2014.02.002.

[38] M. Ż, M. Ruiz, M. Klinkowski, M. Pióro, L. Velasco, Reoptimization of Dynamic Flexgrid Optical Networks After Link Failure Repairs, 7 (2015) 49–61.

[39] B. Chen, J. Zhang, Y. Zhao, C. Lv, W. Zhang, S. Huang, X. Zhang, Optical Fiber Technology Multi-link failure restoration with dynamic load balancing in spectrum-elastic optical path networks, Opt. Fiber Technol. 18 (2012) 21–28. https://doi.org/10.1016/j.yofte.2011.10.002.

[40] Z. Zhu, S. Member, W. Lu, L. Zhang, N. Ansari, Dynamic Service Provisioning in Elastic Optical Networks With Hybrid Single- / Multi-Path Routing, 31 (2013) 15–22.

[41] F. Paolucci, A. Castro, F. Cugini, L. Velasco, P. Castoldi, Multipath restoration and bitrate squeezing in SDN-based elastic optical networks [ Invited ], (2014) 45–57. https://doi.org/10.1007/s11107-014-0444-4.

[42] A. Castro, L. Velasco, M. Ruiz, J. Comellas, Single-path provisioning with multi-path recovery in flexgrid optical networks, Int. Congr. Ultra Mod. Telecommun. Control Syst. Work. (2012) 745–751. https://doi.org/10.1109/ICUMT.2012.6459763.

[43] S. Ramamurthy, L. Sahasrabuddhe, B. Mukherjee, Survivable WDM mesh networks, Light. Technol. J. 21 (2003) 870–883. https://doi.org/10.1109/JLT.2002.806338.

**Table 1** Average value of various parameters of the network for the different protection schemes.

| Network Parameters | ARPANET | | | COST | | |
|---|---|---|---|---|---|---|
| | DPP | SPP | INCB-SPP | DPP | SPP | INCB-SPP |
| BBP | 0.7072 | 0.4041 | 0.3828 | 0.3947 | 0.0455 | 0.0154 |
| RT (µs) | 8715 | 16338 | 4683 | 8201 | 12474 | 3799 |
| BPR | 6.0223 | 2.8712 | 2.7161 | 3.5045 | 1.7549 | 1.7149 |

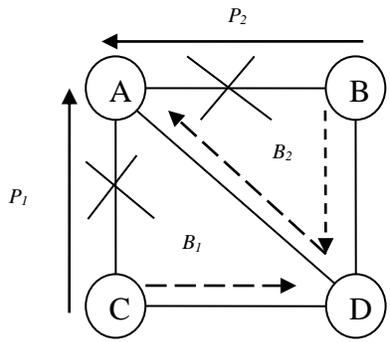 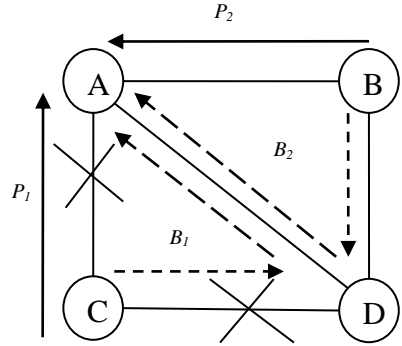

Fig. 1 (a) For Primary ($P_1$) = C-A, $B_1$ (Backup) = C-D-A and For Primary, ($P_2$) = B-A, $B_2$ = B-D-A  (b) For $P_1$= C-A, $B_1$ = C-D-A and for $P_2$ = B-A, $B_2$ =B-D-A.

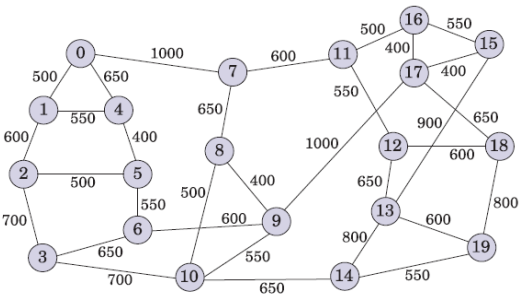 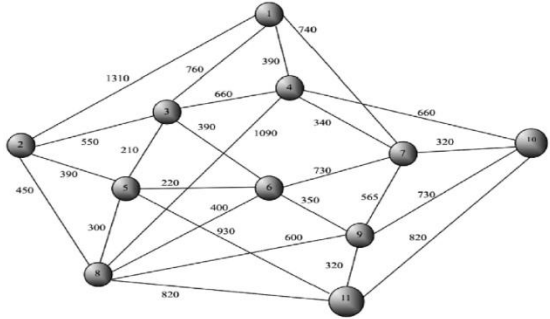

Fig.2 (a) ARPANET topology with 20 nodes and 32 links, (b) COST 239 topology with 11 nodes and 26 links.

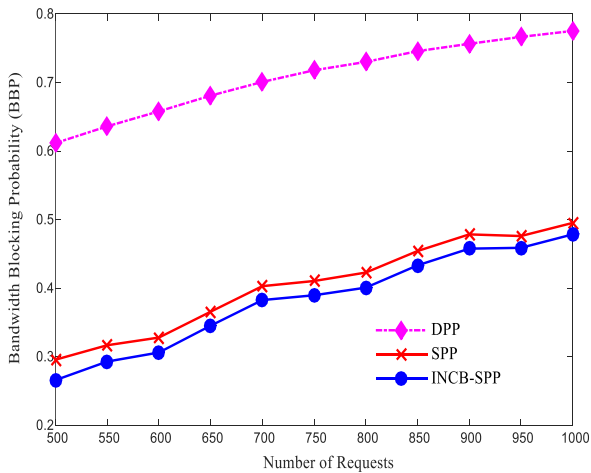 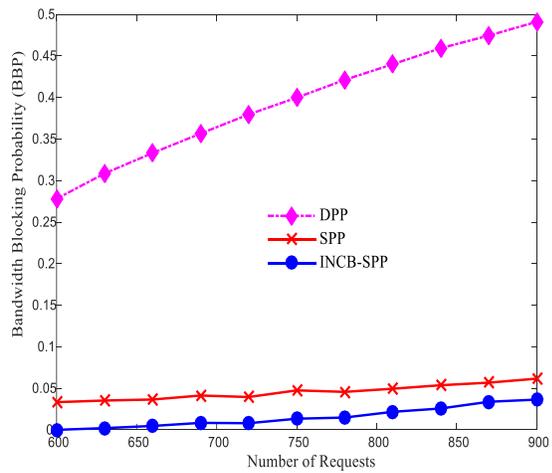

Fig.3 (a) Bandwidth blocking probability (BBP) vs. number of requests for ARPANET (b) BBP vs. number of requests for COST239.

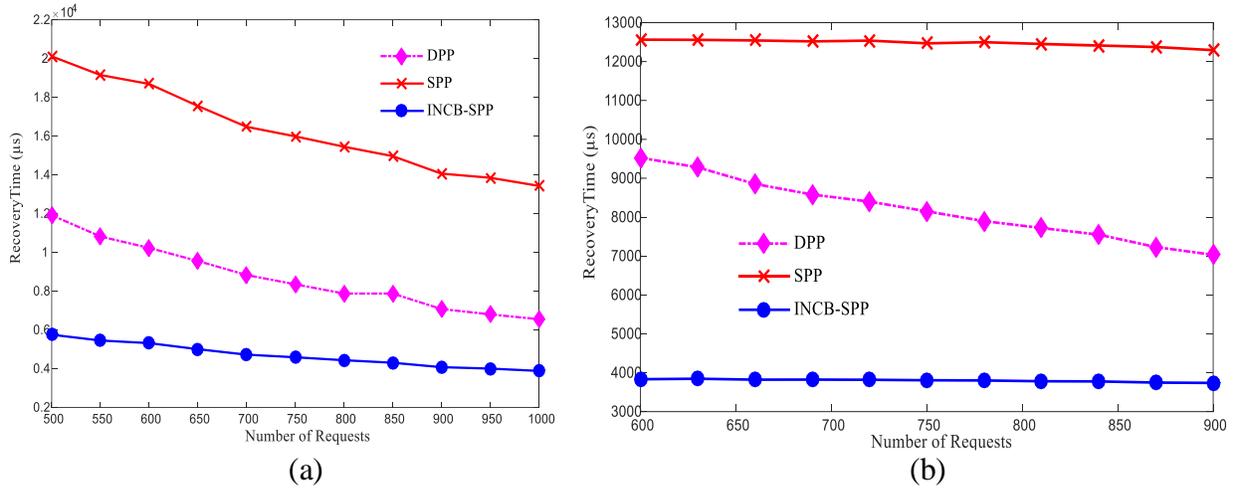

Fig. 4 (a) Recovery time in micro second vs. number of requests for ARPANET (b) recovery time for COST239.

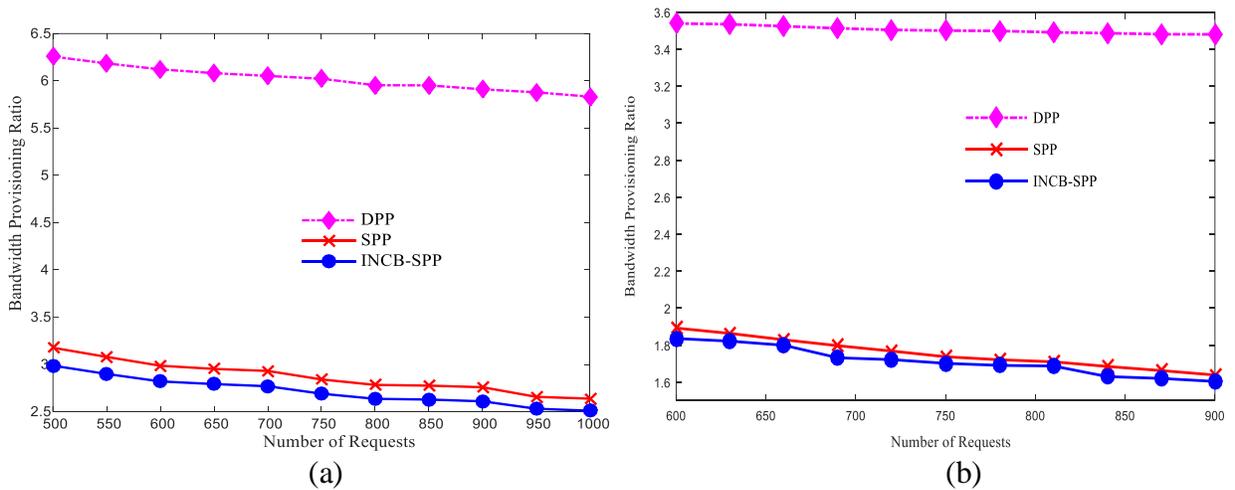

Fig. 5(a) Bandwidth provisioning ratio (BPR) vs. number of requests for ARPANET (b) Bandwidth provisioning ratio vs. number of requests for COST239.